\documentclass[useAMS,usenatbib]{mn2e}
\usepackage{times}
\usepackage{mathptmx}
\usepackage{graphicx}
\usepackage{graphics}
\usepackage{rotating}
\usepackage{amsmath}
\usepackage{amssymb}
\title[GMRT H-ATLAS/GAMA survey]
{A 325-MHz GMRT survey of the {\it Herschel}-ATLAS/GAMA fields}

\author[T. Mauch et al.]
{Tom~Mauch$^{1,2,3}$\thanks{
   E-mail: txmauch@gmail.com},
   Hans-Rainer Kl\"ockner$^{1,4}$, Steve Rawlings$^1$, Matt Jarvis$^{2,5,1}$,
   \newauthor
   Martin J.\ Hardcastle$^2$, Danail Obreschkow$^{1,6}$, D.J.
   Saikia$^{7,8}$ and Mark A.\ Thompson$^2$\\
$^1$Oxford Astrophysics, 
    Denys Wilkinson Building,
    Keble Road, Oxford OX1 3RH\\
$^2$Centre for Astrophysics Research, 
    University of Hertfordshire, 
    College Lane, 
    Hatfield, 
    Hertfordshire AL10 9AB\\
$^3$SKA South Africa, Third Floor, The Park, Park Road, Pinelands, 7405
South Africa\\
$^4$Max-Planck-Institut f\"ur Radioastronomie,
    Auf dem H\"ugel 69,
    53121 Bonn,
    Germany\\
$^5$ Physics Department, University of the Western Cape, Cape Town,
  7535, South Africa\\
$^6$ International Centre for Radio Astronomy Research, University of Western
 Australia, 35 Stirling Highway, Crawley, WA 6009, Australia\\
$^7$ National Centre for Radio Astrophysics, Tata Institute of
 Fundamental Research, Pune University Campus, Ganeshkind P.O., Pune
 411007, India\\
$^8$ Cotton College State University, Panbazar, Guwahati 781001, India\\
}

\date{\today}

\pagerange{\pageref{firstpage}--\pageref{lastpage}}

\pubyear{2013}

\begin{document}

\maketitle

\label{firstpage}

\begin{abstract}

We describe a 325-MHz survey, undertaken with the Giant Metrewave
Radio Telescope (GMRT), which covers a large part of the three
equatorial fields at 9, 12 and 14.5 h of right ascension from the {\it
  Herschel}-Astrophysical Terahertz Large Area Survey (H-ATLAS) in the
area also covered by the Galaxy And Mass Assembly survey (GAMA). The
full dataset, after some observed pointings were removed during the
data reduction process, comprises 212 GMRT pointings covering $\sim90$
deg$^2$ of sky. We have imaged and catalogued the data using a
pipeline that automates the process of flagging, calibration,
self-calibration and source detection for each of the survey
pointings. The resulting images have resolutions of between 14 and 24
arcsec and minimum rms noise (away from bright sources) of $\sim1$ mJy
beam$^{-1}$, and the catalogue contains 5263 sources brighter than
$5\sigma$. We investigate the spectral indices of GMRT sources which
are also detected at 1.4 GHz and find them to agree broadly with
previously published results; there is no evidence for any flattening
of the radio spectral index below $S_{1.4}=10$ mJy. This work adds to
the large amount of available optical and infrared data in the H-ATLAS
equatorial fields and will facilitate further study of the
low-frequency radio properties of star formation and AGN activity in
galaxies out to $z \sim 1$.
\end{abstract}


\section{Introduction}

The \textit{Herschel}-Astrophysical Terahertz Large Area Survey
\citep[H-ATLAS;][]{eales10} is the largest Open Time extragalactic
survey being undertaken with the \textit{Herschel Space Observatory}
\citep{herschel10}. It is a blind survey and aims to provide a wide
and unbiased view of the sub-millimetre Universe at a median redshift
of $1$. H-ATLAS covers $\sim 570$ deg$^2$ of sky at $110$, $160$,
$250$, $350$ and $500$ ${\mu}$m and is observed in parallel mode with
{\it Herschel} using the Photodetector Array Camera
\citep[PACS;][]{pacs} at 110 and 160 ${\mu}$m and the Spectral and
Photometric Imaging Receiver \citep[SPIRE;][]{spire} at 250, 350 and
500 ${\mu}$m. The survey is made up of six fields chosen to have
minimal foreground Galactic dust emission, one field in the northern
hemisphere covering $150$ deg$^2$ (the NGP field), two in the southern
hemisphere covering a total of $250$ deg$^2$ (the SGP fields) and
three fields on the celestial equator each covering $\sim 35$ deg$^2$
and chosen to overlap with the Galaxy and Mass Assembly redshift
survey \citep[GAMA;][]{Driver+11} (the GAMA fields). The H-ATLAS
survey is reaching 5-$\sigma$ sensitivities of (132, 121, 33.5, 37.7,
44.0) mJy at (110, 160, 250, 350, 500) $\mu$m and is expected to
detect $\sim 200,000$ sources when complete \citep{Rigby+11}.

A significant amount of multiwavelength data is available and planned
over the H-ATLAS fields. In particular, the equatorial H-ATLAS/GAMA
fields, which are the subject of this paper, have been imaged in the
optical (to $r \sim 22.1$) as part of the Sloan Digital Sky Survey
\citep[SDSS;][]{sloan} and in the infrared (to $K \sim 20.1$) with the
United Kingdom Infra-Red Telescope (UKIRT) through the UKIRT Infrared
Deep Sky Survey \citep[UKIDSS;][]{ukidss} Large Area Survey (LAS). In
the not-too-distant future, the GAMA fields will be observed
approximately two magnitudes deeper than the SDSS in 4 optical bands
by the Kilo-Degree Survey (KIDS) to be carried out with the Very Large
Telescope (VLT) Survey Telescope (VST), which was the original
motivation for observing these fields. In addition, the GAMA fields
are being observed to $K \sim 1.5-2$ mag. deeper than the level
achieved by UKIDSS as part of the Visible and Infrared Survey
Telescope for Astronomy (VISTA) Kilo-degree Infrared Galaxy (VIKING)
survey, and with the Galaxy Evolution Explorer (GALEX) to a limiting
AB magnitude of $\sim 23$.

In addition to this optical and near-infrared imaging there is also
extensive spectroscopic coverage from many of the recent redshift
surveys. The SDSS survey measured redshifts out to $z\sim0.3$ in the
GAMA and NGP fields for almost all galaxies with $r<17.77$. The
Two-degree Field (2dF) Galaxy Redshift Survey \citep[2dFGRS;][]{2df}
covers much of the GAMA fields for galaxies with $b_{J}<19.6$ and
median redshift of $\sim0.1$. The H-ATLAS fields were chosen to
overlap with the GAMA survey, which is ongoing and aims to measure
redshifts for all galaxies with $r<19.8$ to $z\sim0.5$. Finally, the
WiggleZ Dark Energy survey has measured redshifts of blue galaxies
over nearly half of the H-ATLAS/GAMA fields to a median redshift of
$z\sim0.6$ and detects a significant population of galaxies at
$z\sim1$.

The wide and deep imaging from the far infrared to the ultraviolet and
extensive spectroscopic coverage makes the H-ATLAS/GAMA fields
unparallalled for detailed investigation of the star-forming and AGN
radio source populations. However, the coverage of the H-ATLAS fields
is not quite so extensive in the radio. All of the fields are covered
down to a $5\sigma$ sensitivity of 2.5 mJy beam$^{-1}$ at 1.4 GHz by the National
Radio Astronomy Obervatory (NRAO) Very Large Array (VLA) Sky Survey
\citep[NVSS;][]{nvss}. These surveys are limited by their $\sim45$-arcsec resolution, which makes unambiguous identification of radio
sources with their host galaxy difficult, and by not being deep enough
to find a significant population of star-forming galaxies, which only
begin to dominate the radio-source population below 1 mJy
\citep[e.g.][]{Wilman08}. The Faint Images of the Radio Sky at
Twenty-cm \citep[FIRST;][]{first} survey covers the NGP and GAMA
fields at a resolution of $\sim6$ arcsec down to $\sim0.5$ mJy at 1.4
GHz, is deep enough to probe the bright end of the star-forming galaxy
population, and has good enough resolution to see the morphological
structure of the larger radio-loud AGN, but it must be combined
with the less sensitive NVSS data for sensitivity to extended
structure. Catalogues based on FIRST and NVSS have already been used
in combination with H-ATLAS data to investigate the radio-FIR
correlation \citep{jarvis+10} and to search for evidence for
differences between the star-formation properties of radio galaxies
and their radio-quiet counterparts (\citealt{hardcastle+10,
  hardcastle+12, virdee+13}).

To complement the already existing radio data in the H-ATLAS fields,
and in particular to provide a second radio frequency, we have
observed the GAMA fields (which have the most extensive
multi-wavelength coverage) at 325 MHz with the Giant Metrewave Radio
Telescope \citep[GMRT;][]{gmrtref}. 
The most sensitive GMRT images reach a $1\sigma$
depth of $\sim 1$ mJy beam$^{-1}$ and the best resolution we obtain is $\sim
14$ arcsec, which is well matched to the sensitivity and
resolution of the already existing FIRST data. 
The GMRT data overlaps with the
three $\sim60$-deg$^2$ GAMA fields, and cover a total of
$108$ deg$^2$ in $288$ 15-minute pointings (see Fig.~\ref{noisemaps}).
These GMRT data, used in
conjunction with the available multiwavelength data, will be valuable
in many studies, including an investigation of the radio-infrared
correlation as a function of redshift and as a function of radio
spectral index, the link between star formation and accretion in
radio-loud AGN and how this varies as a function of environment and
dust temperature, and the three-dimensional clustering of radio-source
populations. The data will also bridge the gap between the
well-studied 1.4-GHz radio source populations probed by NVSS and FIRST
and the radio source population below 250 MHz, which will be probed
by the wide area surveys made with the Low Frequency Array
\citep[LOFAR;][]{reflofar} in the coming years.

This paper describes the 325-MHz survey of the H-ATLAS/GAMA regions.
The structure of the paper is as follows. In Section 2 we describe the
GMRT observations and the data. In Section 3 we describe the pipeline
that we have used to reduce the data and in Section 4 we describe the
images and catalogues produced. In Section 5 we discuss the data
quality and in Section 6 we present the spectral index distribution
for the detected sources between 1.4 GHz and 325 MHz. A summary and
prospects for future work are given in Section 7.

\section{GMRT Observations}

\begin{table*}
\centering
\caption{Summary of the GMRT observations.}
\label{obssummary}
\begin{tabular}{lccccc}
\hline
Date & Start Time (IST) & Hours Observerd & N$_{\rm{antennas}}$ & Antennas Down & Comments \\

\hline
2009, Jan 15 & 21:00 & 14.0 & 27 & C01,S03,S04 & C14,C05 stopped at 09:00 \\
2009, Jan 16 & 21:00 & 15.5 & 27 & C01,S02,S04 & C04,C05 stopped at 09:00 \\
2009, Jan 17 & 21:00 & 15.5 & 29 & C01 & C05 stopped at 06:00 \\
2009, Jan 18 & 21:00 & 16.5 & 26 & C04,E02,E03,E04 & C05 stopped at 09:00 \\
2009, Jan 19 & 22:00 & 16.5 & 29 & C04 & \\
2009, Jan 20 & 21:00 & 13.5 & 29 & C01 & 20min power failure at 06:30 \\
2009, Jan 21 & 21:30 & 13.0 & 29 & S03 & Power failure after 06:00 \\
2010, May 17 & 16:00 & 10.0 & 26 & C12,W01,E06,E05 & \\
2010, May 18 & 17:00 & 10.0 & 25 & C11,C12,S04,E05,W01 & E05 stopped at 00:00 \\
2010, May 19 & 18:45 & 10.5 & 25 & C12,E05,C05,E03,E06 & 40min power failure at 22:10 \\
2010, Jun 4  & 13:00 & 12.0 & 28 & W03,W05 & \\
\hline
\end{tabular}
\end{table*}

\subsection{Survey Strategy}

\begin{figure}
\centering
\includegraphics[width=\linewidth]{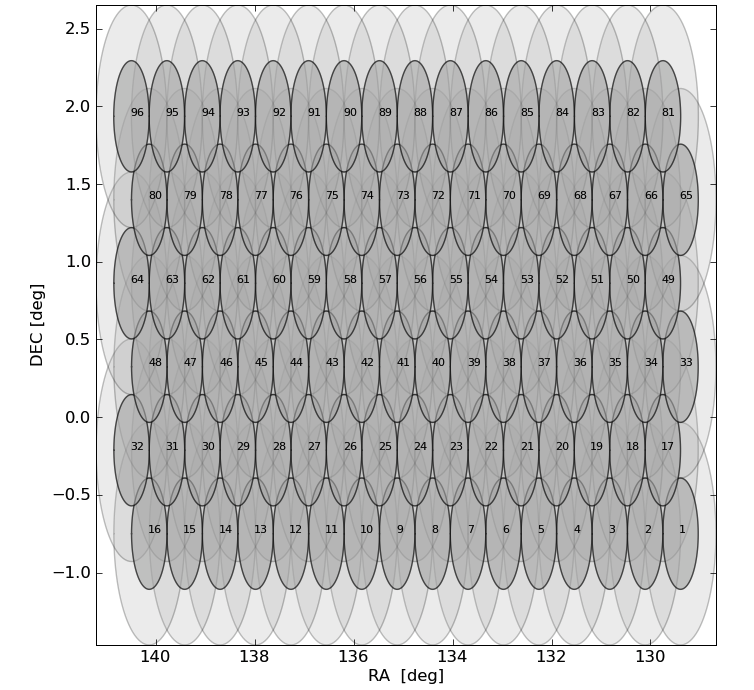}
\caption{The 96 hexagonal GMRT pointings for the 9-h H-ATLAS/GAMA
  fields. The pointing strategy for the 12- and 14.5-h fields is
  similar. The dark grey ellipses (circles on the sky) show the
  42-arcmin region at the centre of each pointing; the light grey
  ellipses (circles) show the 84-arcmin primary beam.}
\label{pointings}
\end{figure}

The H-ATLAS/GAMA regions that have been observed by the \textit{Herschel Space
  Observatory} and are followed up in our GMRT survey are made up of
three separate fields on the celestial equator. The three fields are
centered at 9 h, 12 h, and 14.5 h right ascension
(RA) and each spans approximately 12 deg in RA and 3 deg in
declination to cover a total of 108 deg$^2$ (36 deg$^2$ per field).
The Full Width at Half Maximum (FWHM) of the primary beam of the GMRT
at 325 MHz is 84 arcmin. In order to cover each H-ATLAS/GAMA
field as uniformly and efficiently as possible, we spaced the
pointings in an hexagonal grid separated by 42 arcmin. An example
of our adopted pointing pattern is shown in Fig.~\ref{pointings};
each field is covered by 96 pointings, with 288 pointings in the
complete survey.

\subsection{Observations}

\begin{figure*}
\centering
\includegraphics[width=\textwidth]{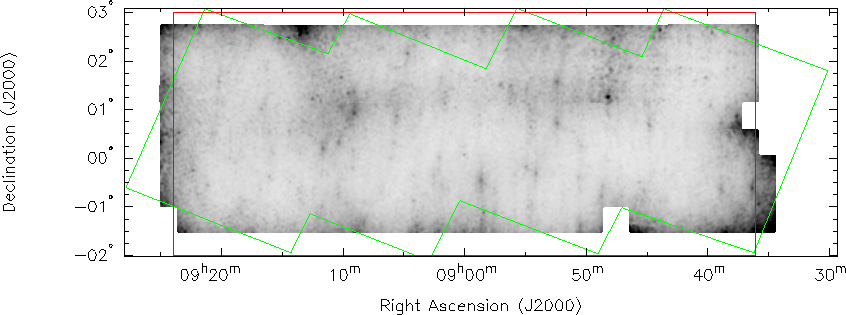}
\includegraphics[width=\textwidth]{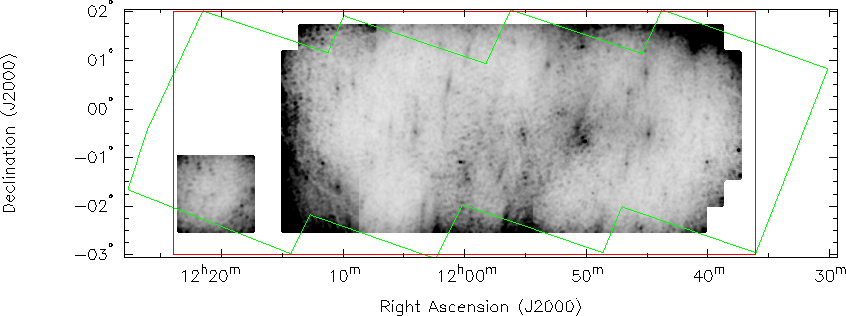}
\includegraphics[width=\textwidth]{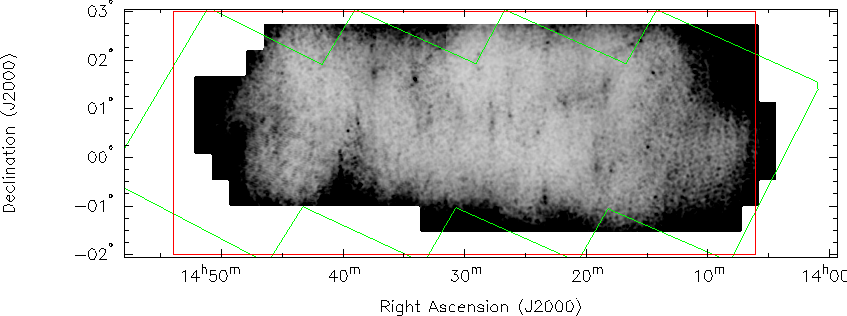}
\caption{Coverage and rms noise maps for the three survey fields. For
  all three fields, the greyscale runs from 0 (white) to 6 mJy
  beam$^{-1}$ (black). White areas were not covered by the survey,
  either intentionally or due to loss of data due to RFI or other
  instrumental problems. The overplotted red boxes show the GAMA survey areas and the
  green lines denote the boundary of the H-ATLAS observations.}
\label{noisemaps}
\end{figure*}

\begin{figure*}
\centering
\includegraphics[width=7cm,angle=-90]{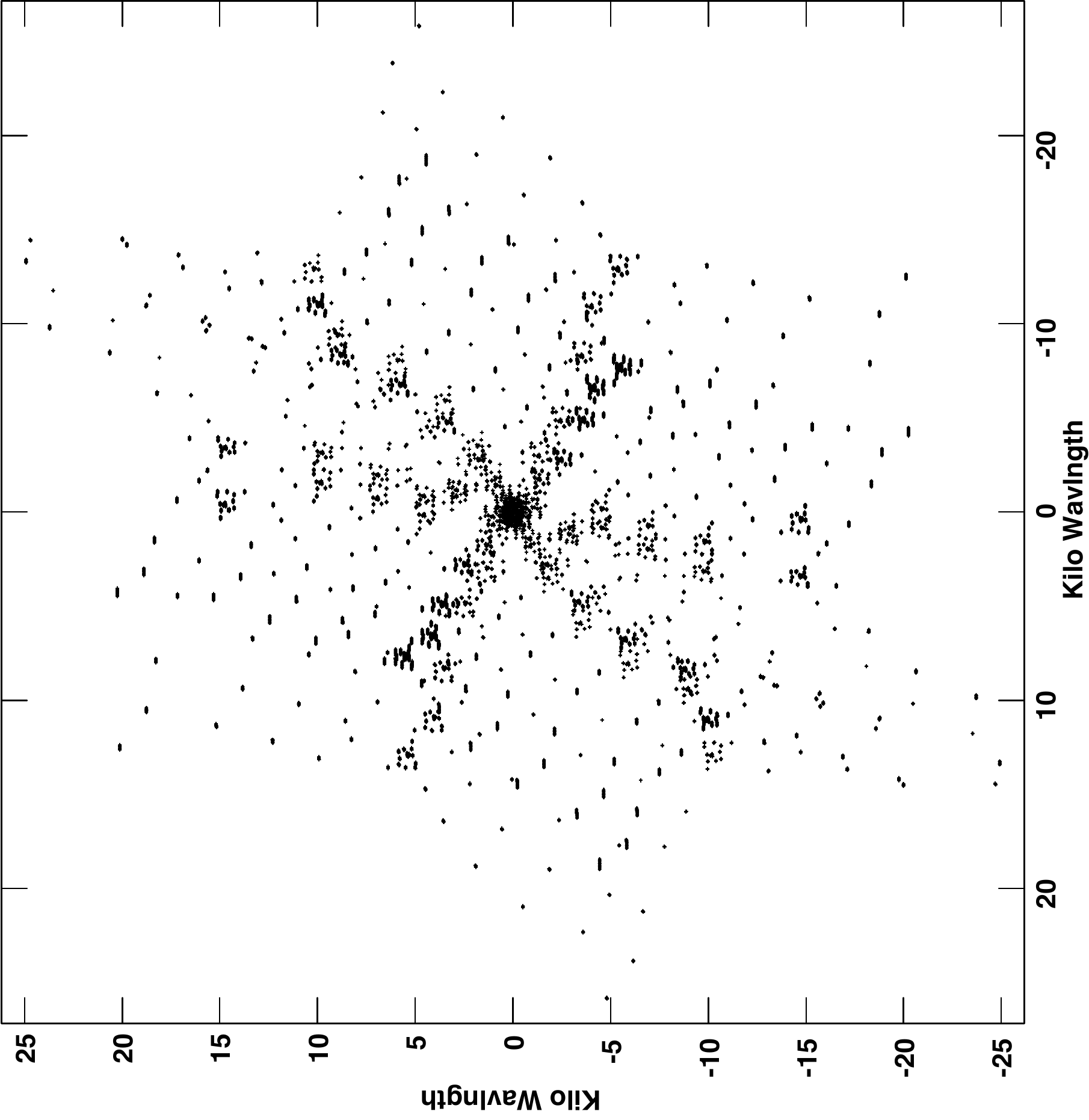}%
\hspace{2cm}
\includegraphics[width=7cm,angle=-90]{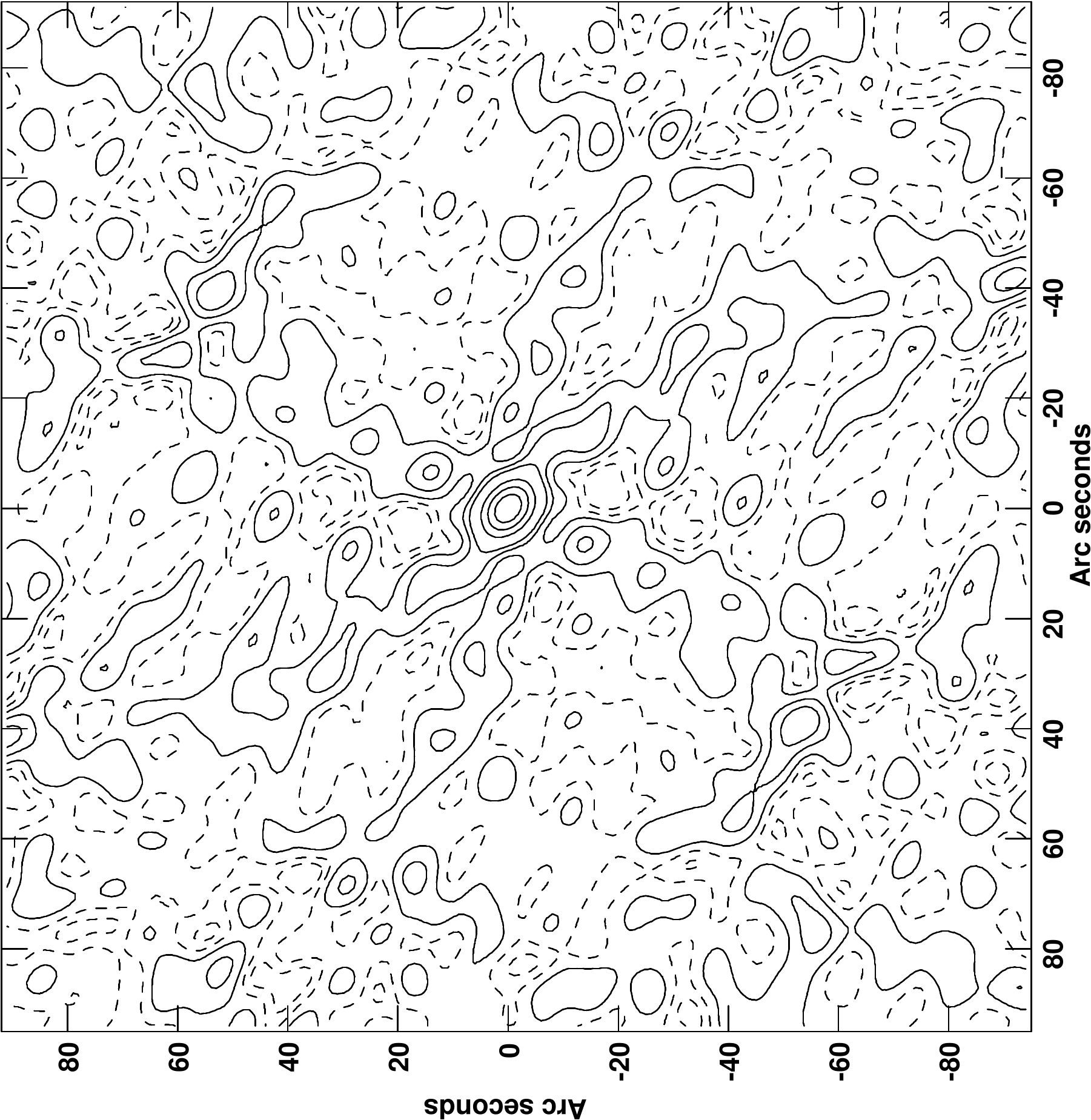}
\caption{\textit{Left:} Typical $uv$ coverage of a survey pointing. This shows the $uv$ coverage
of the pointing PNC71 (see Section \ref{catalogue} for an explanation
of this notation) which had $2\times7.5$ minute scans, separated by 3.5 h. \textit{Right:} The dirty beam
corresponding to the $uv$ coverage on the right, with robust weighting. Contours are plotted at 80, 60, 40, 15, 8, 2, $-$2 and $-$5 
per cent of the peak.}
\label{uvcoverage}
\end{figure*}

Observations were carried out in three runs in Jan 2009 (8 nights) and
in May 2010 (3 nights) and in June 2010 (1 night).
Table~\ref{obssummary} gives an overview of each night's observing. On
each night as many as 5 of the 30 GMRT antennas could be offline for various reasons,
including being painted or problems with the hardware backend. On two
separate occasions (Jan 20 and May 19) power outages at the telescope
required us to stop observing, and on one further occasion on Jan 21 a
power outage affected all the GMRT baselines outside the central
square. Data taken during the Jan 21 power outage were later
discarded.

Each night's observing consisted of a continuous block of 10-14 h
beginning in the early evening or late afternoon and running through
the night. Night-time observations were chosen so as to minimise the
ionopheric variations. We used the GMRT with its default parameters at
325 MHz and its hardware backend (GMRT Hardware Backend; GHB), two
16 MHz sidebands (Upper Sideband; USB, and Lower Sideband; LSB) on
either side of 325 MHz, each with 128 channels, were used. The
integration time was set to 16.7 s.

The flux calibrators 3C147 and 3C286 were observed for 10 minutes at
the beginning and towards the end of each night's oberving. 
We assumed 325-MHz flux densities of 46.07 Jy for 3C\,147 and
24.53 Jy for 3C\,286, using the standard VLA (2010) model provided
by the {\tt AIPS} task {\tt SETJY}. Typically
the observing on each night was divided into 3 $\sim4-5$-h sections,
concentrating on each of the 3 separate fields in order of increasing
RA. The 9-h and 12-h fields were completely covered in the Jan 2009
run and we carried out as many observations of the 14.5-h field as
possible during the remaining nights in May and June 2010. The
resulting coverage of the sky, after data affected by power outages or
other instrumental effects had been taken into account, is shown in
Fig.\ \ref{noisemaps}, together with an indication of the relationship
between our sky coverage and that of GAMA and H-ATLAS.

Each pointing was observed for a total of 15 minutes in two
7.5-min scans, with each scan producing $\sim 26$ records using
the specified integration time. The two scans on each pointing were
always separated by as close to 6 h in hour angle as possible so
as to maximize the $uv$ coverage for each pointing. The $uv$
coverage and the dirty beam of a typical pointing, observed in two 
scans with an hour-angle separation of 3.5 h, is shown in
Fig.~\ref{uvcoverage}.

\subsection{Phase Calibrators}

One phase calibrator near to each field was chosen and was verified to
have stable phases and amplitudes on the first night's observing. All
subsequent observations used the same phase calibrator, and these
calibrators were monitored continuously during the observing to ensure
that their phases and amplitudes remained stable. The positions and flux
densities of the phase calibrators for each field are listed in
Table~\ref{phasecals}. Although there are no 325-MHz observations of
the three phase calibrators in the literature, we estimated their
325-MHz flux densities that are listed in the table using their
measured flux densities from the 365-MHz Texas survey
\citep{texassurvey} and extrapolated to 325 MHz assuming a spectral
index of $\alpha=-0.8$\footnote{$S \propto \nu^\alpha$}.

Each 7.5-minute scan on source was interleaved with a 2.5-minute scan
on the phase calibrator in order to monitor phase and amplitude
fluctuations of the telescope, which could vary significantly during
an evening's observing. During data reduction we discovered that the
phase calibrator for the 14.5-h field (PHC00) was significantly
resolved on scales of $\sim 10$ arcsec. It was therefore necessary to
flag all of the data at $uv$ distance $>20$ k$\lambda$ from
the 14.5-h field. This resulted in degraded resolution and
sensitivity in the 14.5-h field, which will be discussed in later
sections of this paper.

During observing the phases and amplitudes of the phase calibrator
measured on each baseline were monitored. The amplitudes
  typically varied smoothly by $<30$ per cent in amplitude for the
  working long baselines and by $<10$ per cent for the working short
  baselines. We can attribute some of this effect to variations in the
  system temperature, but since the effects are larger on long
  baselines it may be that slight resolution of the calibrators is
  also involved. Phase variations on short to medium baselines were of
  the order of tens of degrees per hour, presumably due to ionospheric
  effects. On several occasions some baselines showed larger phase
and amplitude variations, and these data were discarded during the
data reduction.

\begin{table}
\caption{The phase calibrators for the three fields.}
\label{phasecals}
\begin{tabular}{ccccc}
\hline
Calibrator & Field &RA (J2000) & Dec. (J2000) & $S_{325\,{\rm MHz}}$ \\
Name & & \textit{hh mm ss.ss} & \textit{dd mm ss.ss} & Jy \\
\hline
PHA00  & 9-hr & 08 15 27.81 & -03 08 26.51 & 9.3 \\
PHB00  & 12-hr & 11 41 08.24 & +01 14 17.47 & 6.5 \\
PHC00  & 14.5-hr & 15 12 25.35 & +01 21 08.64 & 6.7 \\
\hline
\end{tabular}
\end{table}

\section{The Data Reduction Pipeline}

The data handling was carried out using an automated calibration
and imaging pipeline. The pipeline is based on {\sc python}, {\sc aips} and
{\sc ParselTongue} (Greisen 1990; Kettenis 2006) and has been specially
developed to handle GMRT data. The pipeline performs a full cycle of
data calibration, including automatic flagging, delay corrections,
absolute amplitude calibration, bandpass calibration, a multi-facet
self-calibration process, cataloguing, and evaluating the final
catalogue. A full description of the GMRT pipeline and the
calibration will be provided elsewhere (Kl\"ockner in prep.).


\subsection{Flagging}
\label{flagging}

The GMRT data varies significantly in quality over time; in particular,
some scans had large variations in amplitude and/or phase over short
time periods, presumably due either to instrumental problems
  or strong ionospheric effects. The phases and amplitudes on each baseline were therefore 
initially inspected manually and any scans with obvious problems were
excluded prior to running the automated flagging procedures. Non-working
antennas listed in Table~\ref{obssummary} were also discarded at this stage.
Finally, the first and last 10 channels of the data were removed as the data
quality was usually poor at the beginning and end of the bandpass. 

After the initial hand-flagging of the most seriously affected data an
automated flagging routine was run on the remaining data. The
automatic flagging checked each scan on each baseline and fitted a 2D
polynomial to the spectrum which was then subtracted from it.
Visibilities $>3\sigma$ from the mean of the background-subtracted
data were then flagged, various kernels were then applied to the data
and also $3\sigma$ clipped and the spectra were gradient-filtered and
flagged to exclude values $>3\sigma$ from the mean. In addition, all
visibilities $>3\sigma$ from the gravitational centre of the
real-imaginary plane were discarded. Finally, after all flags had been
applied any time or channel in the scan which had had $>40$ per cent
of its visibilities flagged was completely removed.

On average, 60 per cent of a night's data was retained after all hand
and automated flagging had been performed. However at times
particularly affected by Radio Frequency Interference (RFI) as little
as 20 per cent of the data might be retained. A few scans ($\sim10$
per cent) were discarded completely due to excessive RFI during their
observation.

\subsection{Calibration and Imaging}
\label{imagepipe}

After automated flagging, delay corrections were determined via the
{\sc aips} task {\tt FRING} and the automated flagging was repeated on
the delay-corrected data. Absolute amplitude calibration was then
performed on the flagged and delay corrected dataset, using the {\sc
  aips} task {\tt SETJY}. The {\sc aips} calibration routine {\tt
  CALIB} was then run on channel 30, which was found to be stable
across all the different night's observing, to determine solutions for
the phase calibrator. The {\sc aips} task {\tt GETJY} was used to
estimate the flux density of the phase-calibrator source (which was
later checked to be consistent with other catalogued flux densities
for this source, as shown in Table~\ref{phasecals}). The bandpass
calibration was then determined using {\tt BPASS} using the
cross-correlation of the phase calibrator. Next, all calibration and
bandpass solutions were applied to the data for the phase calibrator
and the amplitude and phase versus $uv$-distance plots were checked to
ensure the calibration had succeded.

The calibration solutions of the phase-calibrator source were then
applied to the target pointing, and a multi-facet imaging and phase
self-calibration process was carried out in order to increase the
image sensitivity. To account for the contributions of the $w$-term in
the imaging and self-calibration process the field of view was divided
into sub-images; the task {\tt SETFC} was used to produce the facets.
The corrections in phase were determined using a sequence of
decreasing solution intervals starting at 15 min and ending at 3 min
(15, 7, 5, 3). At each self-calibration step a local sky model was
determined by selecting clean components above $5\sigma$ and
performing a model fit of a single Gaussian in the image plane using
{\tt SAD}. The number of clean components used in the first
self-calibration step was 50, and with each self-calibration step the
number of clean components was increased by 100.

After applying the solutions from the self-calibration process the
task {\tt IMAGR} is then used to produce the final sub-images. These
images were then merged into the final image via the task {\tt FLATN},
which combines all facets and performs a primary beam correction. The
parameters used in {\tt FLATN} to account for the contribution of the
primary beam (the scaled coefficients of a polynomial in the off-axis
distance) were: -3.397, 47.192, -30.931, 7.803.

\subsection{Cataloguing}
\label{catadesc}

The LSB and USB images that were produced by the automated imaging
pipeline were subsequently run through a cataloguing routine. As well
as producing source catalogues for the survey, the cataloguing routine
also compared the positions and flux densities measured in each image
with published values from the NVSS and FIRST surveys as a
figure-of-merit for the output of the imaging pipeline. This allowed
the output of the imaging pipeline to be quickly assesed; the
calibration and imaging could subsequently be run with tweaked
parameters if necessary.

The cataloguing procedure first determined a global rms noise
($\sigma_{\rm global}$) in the input image by running {\tt IMEAN} to
fit the noise part of the pixel histogram in the central 50 per cent
of the (non-primary-beam corrected) image. In order to mimimise any
contribtion from source pixels to the calculation of the image rms,
{\tt IMEAN} was run iteratively using the mean and rms measured from
the previous iteration until the measured noise mean changed by less
than 1 per cent.

The limited dynamic range of the GMRT images and errors in calibration
can cause noise peaks close to bright sources to be fitted in a basic
flux-limited cataloguing procedure. We therefore model background
noise variation in the image as follows:

\begin{enumerate}
\item Isolated point sources brighter than $100\sigma_{\rm global}$
  were found using {\tt SAD}. An increase in local source density
  around these bright sources is caused by noise peaks and artefacts
  close to them. Therefore, to determine the area around each bright
  source that has increased noise and artefacts, the source density of
  $3\sigma_{\rm global}$ sources as a function of radius from the
  bright source position was determined. The radius at which the local
  source density is equal to the global source density of all
  $3\sigma_{\rm global}$ sources in the image was then taken as the
  radius of increased noise around bright sources.

\item To model the increased noise around bright sources a
  \textit{local} dynamic range was found by determining the ratio of
  the flux density of each $100\sigma_{\rm global}$ bright source to the
  brightest $3\sigma_{\rm global}$ source within the radius determined
  in step (i). The median value of the local dynamic range for all
  $100\sigma_{\rm global}$ sources in the image was taken to be the
  local dynamic range. This median local dynamic range determination
  prevents moderately bright sources close to the $100\sigma$ source
  from being rejected, which would happen if \textit{all} sources
  within the computed radius close to bright sources were rejected.

\item A local rms ($\sigma_{\rm local}$) map was made from the input
  image using the task {\tt RMSD}. This calculates the rms of pixels
  in a box of $5$ times the major axis width of the restoring beam and
  was computed for each pixel in the input image. {\tt RMSD} iterates
  its rms determination 30 times and the computed histogram is clipped
  at $3\sigma$ on each iteration to remove the contribution of source
  data to the local rms determination.

\item We then added to this local rms map a Gaussian at the position of
  each $100\sigma_{\rm global}$ source, with width determined from the
  radius of the local increased source density from step (i) and peak
  determined from the median local dynamic range from step (ii).

\item A local mean map is constructed in a manner similar to that
  described in step (iii).

\end{enumerate}

Once a local rms and mean model has been produced the input map was
mean-subtracted and divided by the rms model. This image was then run
through the {\tt SAD} task to find the positions and sizes of all
$5\sigma_{\rm local}$ peaks. Eliptical Gaussians were fitted to the
source positions using {\tt JMFIT} (with peak flux density as the only free
parameter) on the original input image to determine the peak and total
flux density of each source. Errors in the final fitted parameters were
determined by summing the equations in \citet{c97} (with $\sigma_{\rm
  local}$ as the rms), adding an estimated $5$ per cent GMRT calibration
uncertanty in quadrature.

Once a final $5\sigma$ catalogue had been produced from the input
image, the sources were compared to positions and flux densities from
known surveys that overlap with the GMRT pointing (i.e., FIRST and NVSS) as a
test of the image quality and the success of the calibration. Any
possible systematic position offset in the catalogue was computed by
comparing the positions of $>15\sigma$ point sources to their
counterparts in the FIRST survey (these are known to be accurate to
better than 0.1 arcsec \citep{first}). For this comparison, a point
source was defined as being one whose fitted size is smaller than the
restoring beam plus 2.33 times the error in the fitted size (98 per
cent confidence), as was done in the NVSS and SUMSS surveys
\citep{nvss,sumss}.

The flux densities of all catalogue sources were compared to the flux
densities of sources from the NVSS survey. At the position of each
NVSS source in the image area, the measured flux densities of each
GMRT source within the NVSS source area were summed and then converted
from 325 MHz to 1.4 GHz assuming a spectral index of $\alpha=-0.7$.
We chose $\alpha=-0.7$ because it is the median spectral index of
radio souces between 843 MHz and 1.4 GHz found between the SUMSS and
NVSS surveys \citep{sumss}; it should therefore serve to indicate whether
any large and systematic offsets can be seen in the distribution of
measured flux densities of the GMRT sources.

\subsection{Mosaicing}
\label{mosaicing}

The images from the upper and lower sidebands of the GMRT that had been
output from the imaging pipeline described in Section~\ref{imagepipe}
were then coadded to produce uniform mosaics. In order to remove the
effects of the increased noise at the edges of each pointing due to
the primary beam and produce a survey as uniform as possible in
sensitivity and resolution across each field, all neighbouring
pointings within 80 arcmin of each pointing were co-added to
produce a mosaic image of $100\times100$ arcmin. This section
describes the mosaicing process in detail, including the combination
of the data from the two sidebands.

\begin{figure}
\centering
\includegraphics[width=\linewidth]{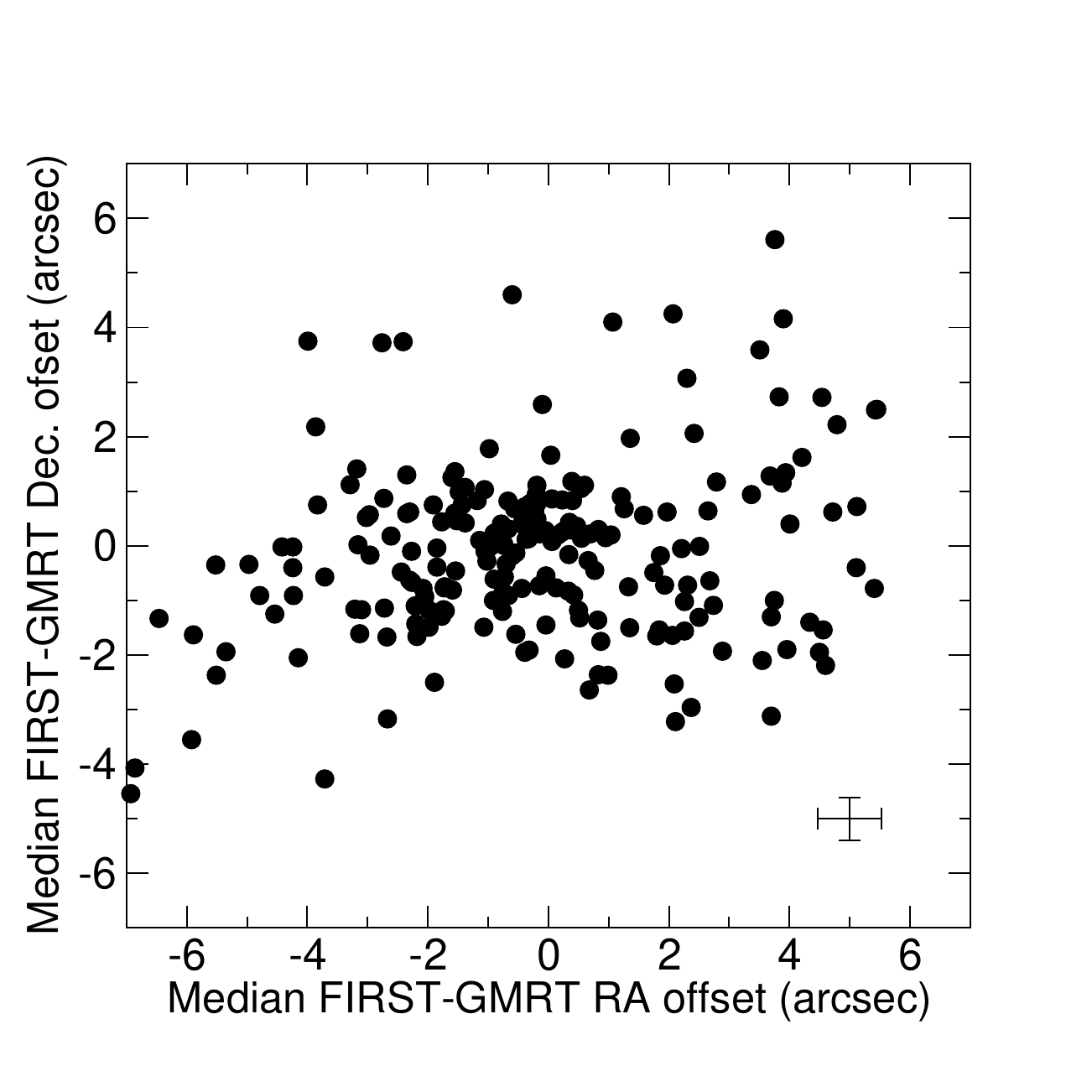}
\caption{The offsets in RA and declination between $15\sigma$
point sources in the GMRT survey that are detected in the FIRST survey. Each
point in the plot is the median offset for all sources in an entire field. The error bars
in the bottom right of the Figure show the rms in RA and declination
from Fig.~\ref{finaloffs}.}
\label{posnoffsets}
\end{figure}

\subsubsection{Combining USB+LSB data}

We were unable to achieve improved signal-to-noise in images produced
by coadding the data from the two GMRT sidebands in the $uv$
plane, so we instead chose to image the USB and LSB data separately
and then subsequently co-add the data in the image plane, which always
produced output images with improved sensitivity. During the process
of co-adding the USB and LSB images, we regridded all of them to a
2 arcsec pixel scale using the {\sc aips} task {\sc regrid}, shifted
the individual images to remove any systematic position offsets, and
smoothed the images to a uniform beam shape across each of the three
survey fields.

Fig.~\ref{posnoffsets} shows the distribution of the median offsets
between the GMRT and FIRST positions of all $>15\sigma$ point sources
in each USB pointing output from the pipeline. The offsets measured
for the LSB were always within 0.5 arcsec of the corresponding USB
pointing. These offsets were calculated for each pointing using the
method described in Section~\ref{catadesc} as part of the standard
pipeline cataloguing routine. As the Figure shows, there was a
significant distribution of non-zero positional offsets between our
images and the FIRST data, which was usually larger than the scatter
in the offsets measured per pointing (shown as an error bar on the
bottom right of the figure). It is likely that these offsets
are caused by ionospheric phase errors, which will largely be refractive
at 325 MHz for the GMRT.
Neighbouring images in the survey can have significantly different
FIRST-GMRT position offsets, and coadding these during the mosaicing
process may result in spurious radio-source structures and flux
densities in the final mosaics. Because of this, the measured offsets
were all removed using the {\sc aips} task {\sc shift} before
producing the final coadded USB+LSB images.

\begin{figure}
\centering
\includegraphics[width=\linewidth]{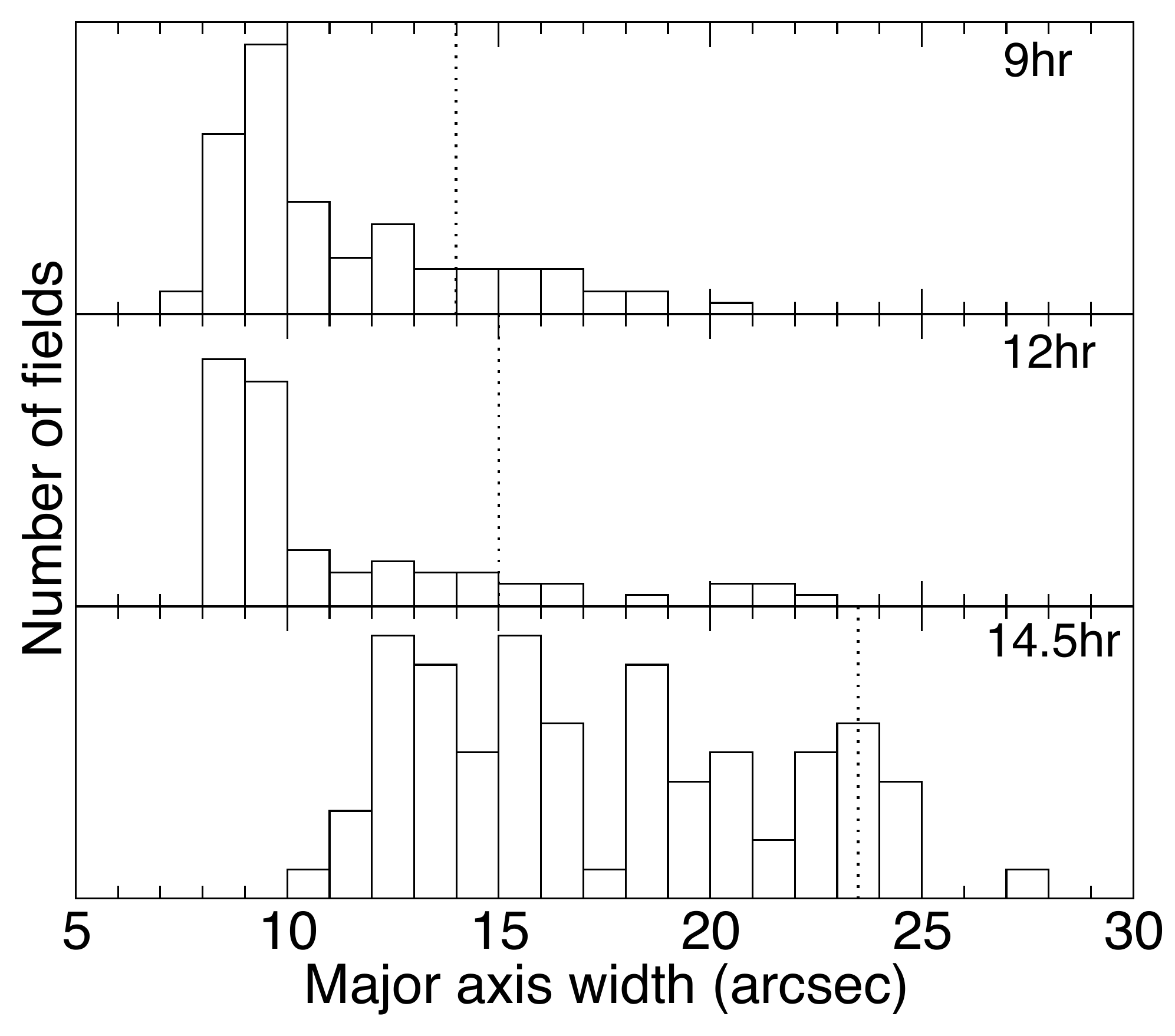}
\caption{The distribution of the raw clean beam major axis FWHM in
  each of the three H-ATLAS/GAMA fields from the USB+LSB images output from
  the imaging pipeline. The dotted line shows the width of the
  convolving beam used before the mosaicing process. Images with raw
  clean beam larger than our adopted cutoffs have been discarded from
  the final dataset.}
\label{beamsizes}
\end{figure}

Next, the USB+LSB images were convolved to the same resolution before
they were co-added; the convolution minimises artefacts resulting from
different source structures at different resolution, and in any case
is required to allow flux densities to be measured from the resulting co-added
maps. Fig.~\ref{beamsizes} shows the distribution in restoring beam
major axes in the images output from the GMRT pipeline. The beam minor
axis was always better than 12 arcsec in the three surveyed fields.
In the 9-h and 12-h fields, the majority of images had better
than 10-arcsec resolution. However, roughly 10 per cent of them are
significantly worse; this can happen for various reasons but is mainly
caused by the poor $uv$ coverage produced by the
$2\times7.5$-minute scans on each pointing. Often, due to scheduling
constraints, the scans were observed immediately after one another
rather than separated by 6 h which can limit the distribution of
visibilities in the $uv$ plane. In addition, when even a few of
the longer baselines are flagged due to interference or have problems
during their calibration, the resulting image resolution can be
degraded.

The distribution of restoring beam major axes in the 14.5-h field is
much broader. This is because of the problems with the phase
calibrator outlined in Section~\ref{phasecals}. All visibilities in
excess of $20$ k$\lambda$ were removed during calibration of the
14.5-h field and this resulted in degraded image resolution.

The dotted lines in Fig.~\ref{beamsizes} show the width of the
beam used to convolve the images for each of the fields before coadding USB+LSB images. We have used a
resolution of 14 arcsec for the 9-h field, 15 arcsec for the
12-h field and 23.5 arcsec for the 14.5-h field. Images with
lower resolution than these were discarded from the final data at this
stage. Individual USB and LSB images output from the self-calibration
step of the pipeline were smoothed to a circular beam using the
{\sc aips} task {\sc convl}.

After smoothing, regridding and shifting the USB+LSB images, they are
combined after being weighted by their indivdual variances, which were
computed from the square of the local rms image measured during the
cataloguing process. The combined USB+LSB images have all pixels
within 30 arcsec of their edge blanked in order to remove any
residual edge effect from the regridding, position shifting and
smoothing process.

\subsubsection{Producing the final mosaics}

The combined USB+LSB images were then combined with all neighbouring
coadded USB+LSB images within 80 arcmin of their pointing center.
This removes the effects at the edges of the individual pointings
caused by the primary beam correction and improves image sensitivity
in the overlap regions. The final data product consists of one 
combined mosaic for each original GMRT pointing, and therefore the user 
should note that there is significant overlap between each mosaic image.

Each combined mosaic image has a width of $100\times100$ arcmin and
2 arcsec pixels. They were produced from all neighboring images with
pointing centers within 80 arcmin. Each of these individual image was
then regridded onto the pixels of the output mosaic. The {\sc aips}
task {\sc rmsd} was run in the same way described during the
cataloging (i.e. with a box size of 5 times the major axis of the
smoothed beam) on the regridded images to produce local rms noise
maps. The noise maps were smoothed with a Gaussian with a FWHM of
3 arcmin to remove any small-scale variation in them. These smoothed
noise maps were then used to create variance weight maps (from the sum
of the squares of the individual noise maps) which were then in turn
multiplied by each regridded input image. Finally, the weighted input
images were added together.

The final source catalogue for each pointing was produced as described
above from the fully weighted and mosaiced images.

\section{Data Products}

The primary data products from the GMRT survey are a set of FITS
images (one for each GMRT pointing that has not been discarded during
the pipeline reduction process) overlapping the H-ATLAS/GAMA fields,
the $5\sigma$ source catalogues and a list of the image central 
positions.\footnote{Data products are available on line at
  http://gmrt-gama.extragalactic.info .} This section briefly describes the imaging data and the 
format of the full catalogues.

\subsection{Images}

\begin{figure*}
\centering
\includegraphics[width=\textwidth]{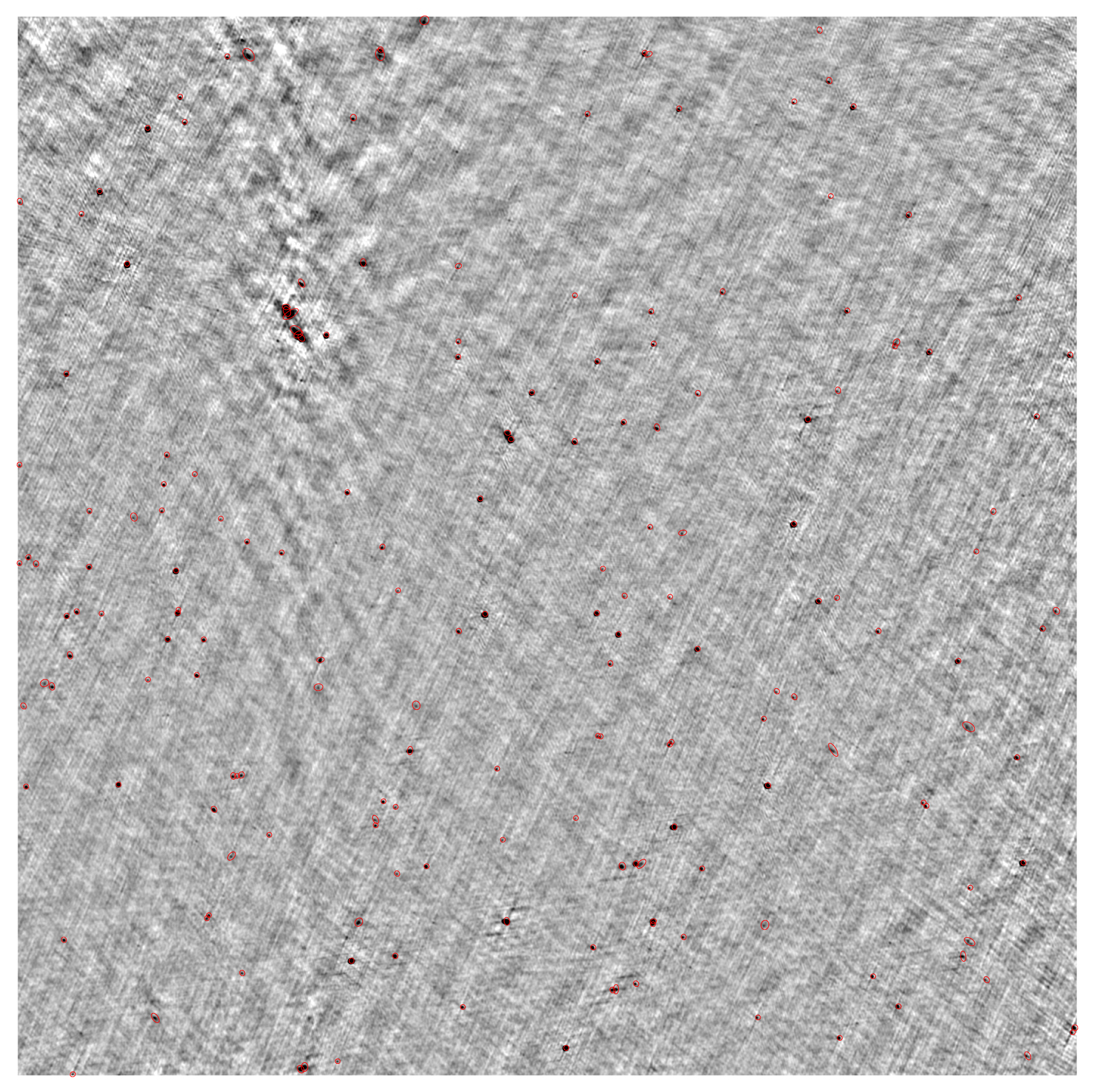}
\caption{An example $100\times100$ arcmin$^2$ mosaic image from the
  survey (pointing PNA35 from the 9-h field, centred on RA 08:43:17.6,
  DEC 00:19:32; see Section \ref{catalogue} for a discussion of the
  field naming scheme) with $5\sigma$ 
catalogued sources circled in red. Note the reduced dynamic range
around the bright (0.9 Jy) double source to the NE.}
\label{eximage}
\end{figure*}

An example of a uniform mosaic image output from the full pipeline is
shown in Fig.~\ref{eximage}.

In each field some of the 96 originally observed pointings had to be
discarded for various reasons that have been outlined in the previous
sections. The full released data set
comprises 80 pointings in the 9-h field, 61 pointings in the
12-h field and 71 pointings in the 14.5-h field. In total 76 out
of the 288 original pointings were rejected. In roughly 50 per
cent of cases they were rejected because of the cutoff in beam
size shown in Fig.~\ref{beamsizes}, while in the other 50 per cent of
cases the $2\times7.5$-minute scans of the pointing were
completely flagged due to interference or other problems with the GMRT
during observing.  The full imaging dataset from the survey
comprises a set of mosaics like the one pictured in
Fig.~\ref{eximage}, one for each of the non-rejected pointings.

\subsection{Catalogue}
\label{catalogue}

\begin{table*}
\small
\caption{Ten example lines from the catalogue; full descriptions of
  each column are in the text.}
\label{catexample}
\scriptsize
\begin{tabular}{cccccccccccccccccc}
\hline
(1)&(2)&(3)&(4)&(5)&(6)&(7)&(8)&(9)&(10)&(11)&(12)&(13)&(14)&(15)&(16)&(17)&(18)\\
RA & Dec. & RA & Dec. & $\Delta$RA & $\Delta$Dec. & $A$ & ${\Delta}A$ &
$S$ & ${\Delta}S$ & Maj & Min & PA & $\Delta$Maj & $\Delta$Min & PA & Local $\sigma$ & Pointing \\ 
\multicolumn{2}{c}{Degrees (J2000)} & $hh$ $mm$ $ss$ & $dd$ $mm$ $ss$ & \multicolumn{2}{c}{arcsec} & \multicolumn{2}{c}{mJy/bm} & \multicolumn{2}{c}{mJy} & \multicolumn{2}{c}{arcsec} & $^\circ$ & \multicolumn{2}{c}{arcsec} & $^\circ$ & mJy/bm & \\
\hline
130.87617 & -00.22886 & 08 43 30.28 & -00 13 43.9 & 2.5 & 1.3 & 6.2 &
1.1 & 15.0 & 3.8 & ---- & ---- & ----- & ---- & ---- & -- & 1.1 &
PNA02 \\
130.87746 & +02.11494 & 08 43 30.59 & +02 06 53.8 & 2.3 & 1.6 & 12.1 &
1.3 & 41.2 & 5.6 & ---- & ---- & ----- & ---- & ---- & -- & 1.4 &
PNA67 \\
130.88025 & +00.48630 & 08 43 31.26 & +00 29 10.7 & 0.5 & 0.3 & 129.7
& 4.7 & 153.6 & 6.7 & 15.9 & 14.6 & 13.8 & 0.5 & 0.4 & 9 & 2.6 & PNA51
\\
130.88525 & +00.49582 & 08 43 32.46 & +00 29 45.0 & 0.5 & 0.4 & 122.2
& 4.6 & 152.1 & 7.0 & ---- & ---- & ----- & ---- & ---- & -- & 2.6 &
PNA51 \\
130.88671 & -00.24776 & 08 43 32.81 & -00 14 51.9 & 0.5 & 0.3 & 106.1
& 3.3 & 171.3 & 5.3 & 21.1 & 15.0 & 80.8 & 0.5 & 0.4 & 1 & 1.0 & PNA02
\\
130.88817 & -00.89953 & 08 43 33.16 & -00 53 58.3 & 0.6 & 0.3 & 59.6 &
2.2 & 118.8 & 5.0 & 26.4 & 14.8 & 87.5 & 0.9 & 0.6 & 1 & 1.2 & PNA03
\\
130.89171 & -00.24660 & 08 43 34.01 & -00 14 47.8 & 0.5 & 0.4 & 34.6 &
1.5 & 38.5 & 2.2 & ---- & ---- & ----- & ---- & ---- & -- & 1.0 &
PNA02 \\
130.89279 & -00.12352 & 08 43 34.27 & -00 07 24.7 & 1.2 & 1.1 & 4.6 &
0.8 & 4.7 & 1.5 & ---- & ---- & ----- & ---- & ---- & -- & 0.8 & PNA35
\\
130.89971 & -00.91813 & 08 43 35.93 & -00 55 05.3 & 2.4 & 1.0 & 7.9 &
1.5 & 13.9 & 3.9 & ---- & ---- & ----- & ---- & ---- & -- & 1.4 &
PNA03 \\
130.90150 & -00.01532 & 08 43 36.36 & -00 00 55.1 & 0.9 & 0.8 & 6.6 &
0.8 & 6.6 & 1.4 & ---- & ---- & ----- & ---- & ---- & -- & 0.8 & PNA03
\\
\hline
\end{tabular}
\end{table*}

Final catalogues were produced from the mosaiced images using the
catalogue procedure described in Section~\ref{catadesc}. The
catalogues from each mosaic image were then combined into 3 full
catalogues covering each of the 9-h, 12-h, and 14.5-h fields. The
mosaic images overlap by about 60 per cent in both RA and declination,
so duplicate sources in the full list were removed by finding all
matches within 15 arcsec of each other and selecting the duplicate
source with the lowest local rms ($\sigma_{\rm local}$) from the full
catalogue; this ensures that the catalogue is based on the best
available image of each source. Removing duplicates reduced the total
size of the full catalogue by about 75 per cent due to the amount of
overlap between the final mosaics.

The resulting full catalogues contain 5263 sources brighter than the
local $5\sigma$ limit. 2628 of these are
in the 9-h field, 1620 in the 12-h field and 1015 in the 14.5-h field.
Table~\ref{catexample} shows 10 random lines of the output catalogue sorted 
by RA. A short description of each of the columns of the 
catalogue follows:

Columns (1) and (2): The J2000 RA and declination of the source
in decimal degrees (the examples given in Table~\ref{catexample} have
reduced precision for layout reasons).

Columns (3) and (4): The J2000 RA and declination of the source
in sexagesimal coordinates.

Columns (5) and (6): The errors in the quoted RA and declination
in arcsec. This is calulated from the quadratic sum of the 
calibration uncertainty, described in Section~\ref{positioncal},
and the fitting uncertainty, calculated
using the equations given by \citet{c97}.

Columns (7) and (8): The fitted peak brightness in units of
mJy beam$^{-1}$ and its associated uncertainty, calculated from the
quadratic sum of the fitting uncertainty from the equations given by
\citet{c97} and the estimated 5 per cent flux calibration uncertainty
of the GMRT. The raw brightness measured from the image has been increased
by 0.9 mJy beam$^{-1}$ to account for the effects of clean bias (see Section~\ref{sec:flux}).

Columns (9) and (10): The total flux density of the source in mJy and its uncertainty
calculated from equations given by \citet{c97}. This equals the fitted
peak brightness if the source is unresolved.

Columns (11), (12) and (13): The major axis FWHM (in arcsec), minor
axis FWHM (in arcsec) and position angle (in degrees east of north) of
the fitted elliptical Gaussian. The position angle is only meaningful
for sources that are resolved (i.e. when the fitted Gaussian is larger
than the restoring beam for the relevant field). As discussed in
Section~\ref{sec:sizes}, fitted sizes are only quoted for sources
that are moderately resolved in their minor axis.

Columns (14), (15) and (16): The fitting uncertanties in the size parameters
of the fitted elliptical Gaussian calculated using equations from
\citet{c97}.

Column (17): The \textit{local} rms noise ($\sigma_{\rm local}$) in mJy beam$^{-1}$
at the source position calculated as described in Section~\ref{catadesc}.
The \textit{local} rms is used to determine the source signal-to-noise
ratio, which is used to determine fitting uncertainties.

Column (18): The name of the GMRT mosaic image containing the source.
These names consist of the letters PN, a letter A, B or C indicating
the 9-, 12- or 14.5-h fields respectively, and a number between 01 and
96 which gives the pointing number within that field (see
Fig.\ \ref{pointings}).

\section{Data Quality}

The quality of the data over the three fields varies considerably due
in part to the different phase and flux calibration sources used for
each field, and also due to the variable observing conditions over the
different nights' observing. In particular on each night's observing,
the data taken in the first half of the night seemed to be much more
stable than that taken in the second half/early mornings. Some power
outages at the telescope contributed to this as well as the variation
in the ionosphere, particularly at sunrise. Furthermore, as described
in Section~\ref{mosaicing}, the poor phase calibrator in the 14.5-h
field has resulted in degraded resolution and sensitivity.

\subsection{Image noise}
\label{sec:noise}

\begin{figure}
\centering
\includegraphics[width=\linewidth]{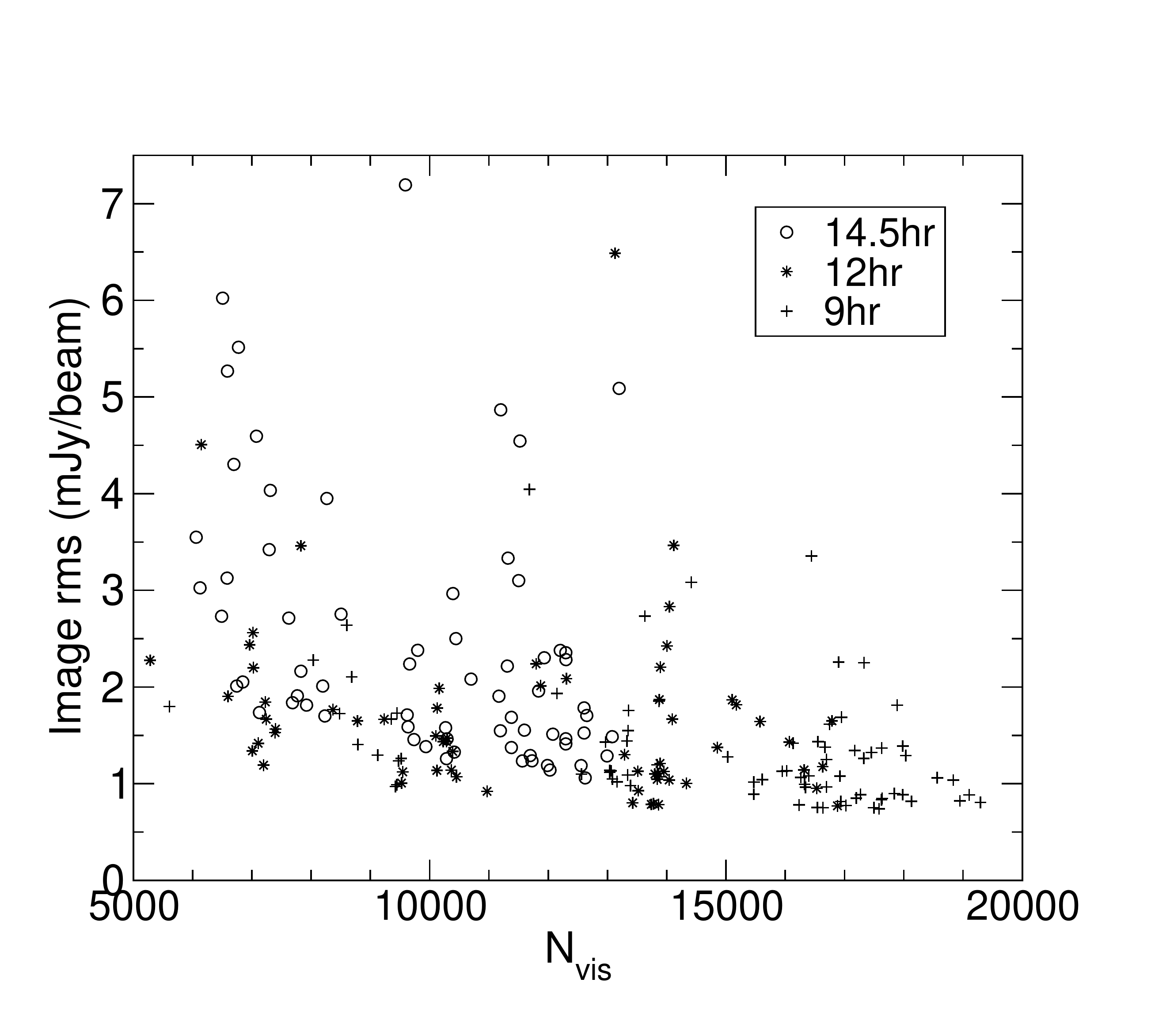}
\caption{The rms noise measured in the central 1000 pixels of each image
plotted against the square root of the number of visibilities. Outliers
from the locus are produced by the increased noise in images around sources brighter than 1 Jy.}
\label{rmsnvis}
\end{figure}

Fig.~\ref{rmsnvis} shows the distribution of the rms noise
measured within a radius of 1000 pixels in the individual GMRT images
immediately after the self-calibration stage of the pipeline, plotted against the
number of visibilities that have contributed to the final image (this
can be seen as a proxy for the effective integration time after flagging). The rms in the
individual fields varies from $\sim 1$ mJy beam$^{-1}$ in those images with the most
visibilities to $\sim 7$ mJy beam$^{-1}$ in the worst case, with the expected trend
toward higher rms noise with decreasing number of visibilities. The
scatter to higher rms from the locus is caused by residual problems in
the calibration and the presence of bright sources in the primary beam
of the reduced images, which can increase the image noise in their
vicinity due to the limited dynamic range of the GMRT observations
($\sim1000:1$). A bright 7 Jy source in the 12-h field and a 5 Jy
source in the 14.5-h field have both contributed to the generally
increased rms noise measured from some images. On average, the most
visibilities have been flagged from the 14.5-h field because of the
restriction we imposed on the $uv$ range of the data. This has also resulted in
higher average noise in the 14.5-h fields.

Fig.~\ref{noisemaps} shows the rms noise maps covering all of the 3
fields. These have been made by averaging the background rms images
produced during the cataloguing of the the final mosaiced images and
smoothing the final image with a Gaussian with a FWHM of 3 arcmin to
remove edge effects between the individual backgound images. The rms
in the final survey is significantly lower than that measured from the
individual images output from the pipeline self-calibration process,
which is a consequence of the large amount of overlap between the
individual GMRT pointings in our survey strategy (see
Fig.~\ref{pointings}). The background rms is $\sim0.6-0.8$ mJy beam$^{-1}$ in
the 9-h field, $\sim0.8-1.0$ mJy beam$^{-1}$ in the 12-h field and
$\sim1.5-2.0$ mJy beam$^{-1}$ in the 14.5-h field. Gaps in the coverage are
caused by having discarded some pointings in the survey due to power
outages at the GMRT, due to discarding scans during flagging as
described in Section~\ref{flagging}, and as a result of pointings
whose restoring beam was larger than the smoothing width during the
mosaicing process (Section~\ref{mosaicing}).

\subsection{Flux Densities}
\label{sec:flux}

The $2\times7.5$-min observations of the GMRT survey 
sample the $uv$ plane sparsely (see Fig.~\ref{uvcoverage}), with long
radial arms which cause the dirty beam to have large radial sidelobes.
These radial sidelobes can be difficult to clean properly during imaging
and clean components which can be subtracted at their position when cleaning
close to the noise can cause the average flux density of all point sources in
the restored image to be systematically reduced. This ``clean bias'' is common
in ``snapshot'' radio surveys and for example was found in the FIRST and NVSS
surveys \citep{first,nvss}.

We have checked for the presence of clean bias in the GMRT data by inserting
500 point sources into the calibrated $uv$ data at random positions and
the re-imaging the modified data with the same parameters as the original
pipeline. We find an average difference between the imaged and input
peak flux densities of $\Delta S_{\rm peak}=-0.9$ mJy beam$^{-1}$ with no
significant difference between the 9hr, 12hr and 14hr fields. A constant
offset of $0.9$ mJy beam$^{-1}$ has been added to the peak flux densities of all
sources in the published catalogues.

As a consistency check for the flux density scale of the survey we can
compare the measured flux densities of the phase calibrator source
with those listed in Table~\ref{phasecals}. The phase calibrator is
imaged using the standard imaging pipeline and its flux density is
measured using {\sc sad} in {\sc aips}. The scatter in the measurments
of each phase calibrator over the observing period gives a measure of
the accuracy of the flux calibration in the survey. In the 9-h
field, the average measured flux density of the phase calibrator PHA00
is 9.5 Jy with rms scatter 0.5 Jy; in the 12-h field, the average
measured flux density of PHB00 is 6.8 Jy with
rms 0.4 Jy; and in the 14.5-h field the average measured flux density
of PHC00 is 6.3 Jy with rms 0.5 Jy. This implies that the
flux density scale of the survey is accurate to within $\sim5$ per
cent; there is no evidence for any systematic offset in the flux
scales.

As there are no other 325-MHz data available for the region covered by
the GMRT survey, it is difficult to provide any reliable external measure of
the absolute quality of the flux calibration. An additional check is provided
by a comparison of the spectral index distribution of sources detected
in both our survey and the 1.4-GHz NVSS survey. We discuss this
comparison further in Section~\ref{sindexsec}.

\subsection{Positions}
\label{positioncal}

\begin{table}
\caption{Median and rms of position offsets between the GMRT and FIRST catalogues.}
\label{offsetdata}
\centering
\begin{tabular}{lcccc}
\hline
Field &  \multicolumn{2}{c}{RA offset (arcsec)} & \multicolumn{2}{c}{Dec. offset (arcsec)} \\
      & median & rms & median & rms \\
\hline
9-h & $-0.04$ & $0.52$ & $0.01$ & $0.31$ \\
12-h & $-0.06$ & $0.54$ & $0.01$ & $0.39$ \\
14.5-h & $0.30$ & $0.72$ & $0.26$ & $0.54$ \\
\hline
\end{tabular}
\end{table}     

\begin{figure}
\centering
\includegraphics[width=\linewidth]{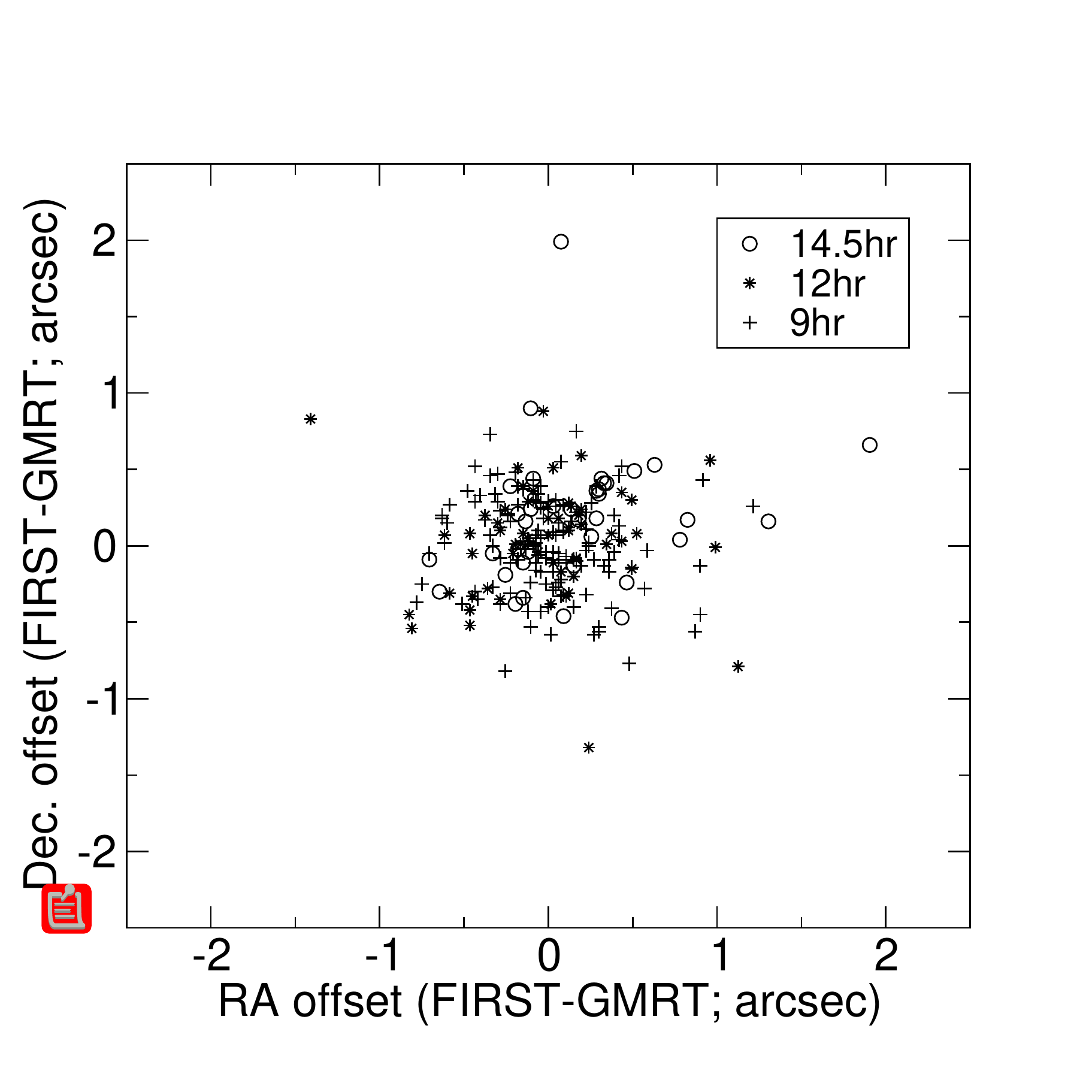}
\caption{The offsets in RA and declination between $>15\sigma$ point
  sources from the GMRT survey that are detected in the FIRST survey.
  The mean offsets in each pointing shown in Fig.~\ref{posnoffsets}
  have been removed. Different point styles are used to denote the
  three different H-ATLAS/GAMA fields to show the effect of the variation in
  the resolution of the GMRT data.}
\label{finaloffs}
\end{figure}

In order to measure the poitional accuracy of the survey, we have
compared the postions of $>15\sigma$ GMRT point sources with sources
from the FIRST survey. Bright point sources in FIRST are known to have
positional accuracy of better than 0.1 arcsec in RA and declination
\citep{first}. We select point sources using the method outlined in
Section~\ref{catadesc}. Postions are taken from the final GMRT source
catalogue, which have had the shifts described in
Section~\ref{mosaicing} removed; the scatter in the measured shifted
positions is our means of estimating the calibration accuracy of the
positions.

Fig.~\ref{finaloffs} shows the offsets in RA and declination between
the GMRT catalogue and the FIRST survey and Table~\ref{offsetdata}
summarizes the mean offsets and their scatter in the three separate
fields. As expected, the mean offset is close to zero in each case,
which indicates that the initial image shifts have been correctly
applied and that no additional position offsets have appeared in the final
mosaicing and cataloguing process. The scatter in the offsets is
smallest in the 9-h field and largest in the 14.5-h field, which
is due to the increasing size of the restoring beam. The rms of the
offsets listed in Table~\ref{offsetdata} give a measure of the
positional calibration uncertainty of the GMRT data; these have been
added in quadrature to the fitting error to produce the errors listed
in the final catalogues.

\subsection{Source Sizes}\label{sec:sizes}

The strong sidelobes in the dirty beam shown in Fig.~\ref{uvcoverage}
extend radially at position angles (PAs) of $40^{\circ}$, $70^{\circ}$
and $140^{\circ}$ and can be as high as 15 per cent of the central
peak up to 1 arcmin from it. Improper cleaning of these sidelobes can
leave residual radial patterns with a similar structure to the dirty
beam in the resulting images. Residual peaks in the dirty beam pattern
can also be cleaned (see the discussion of ``clean bias'' in
Section~\ref{sec:flux}) and this has the effect of enhancing positive
and negative peaks in the dirty beam sidelobes, and leaving an imprint
of the dirty beam structure in the cleaned images. This effect,
coupled with the alternating pattern of positive and negative peaks in
the dirty beam structure (see Fig.~\ref{uvcoverage}), causes sources
to appear on ridges of positive flux squeezed between two negative
valleys. Therefore, when fitting elliptical Gaussians to even
moderately strong sources in the survey these can appear spuriously
extended in the direction of the ridge and narrow in the direction of
the valleys.

These effects are noticeable in our GMRT images (see, for example,
Fig.~\ref{eximage}) and in the distribution of fitted position angles
of sources that appear unresolved in their minor axes (ie. $\phi_{\rm
  min}-\theta_{\rm min} < \sigma_{\rm min}$; where $\phi_{\rm min}$ is
the fitted minor axis size, $\theta_{\rm min}$ is the beam minor axis
size and $\sigma_{\rm min}$ is the rms fitting error in the fitted
minor axis size) and are moderately resolved in their major axes (ie.
$\phi_{\rm maj}-\theta_{\rm maj} > 2\sigma_{\rm maj}$; defined by
analogy with above) from the catalogue. These PAs are clustered on
average at $65^\circ$ in the 9-hr field, $140^\circ$ in the 12-hr
field and at $130^\circ$ in the 14.5-hr field, coincident with the PAs
of the radial sidelobes in the dirty beam shown in
Fig.~\ref{uvcoverage}. The fitted PAs of sources that show some
resolution in their minor axes (ie. $\phi_{\rm min}-\theta_{\rm min} >
\sigma_{\rm min}$) are randomly distributed between $0^\circ$ and
$180^\circ$ as is expected of the radio source population. We
therefore only quote fitted source sizes and position angles for
sources with $\phi_{\rm min}-\theta_{\rm min} > \sigma_{\rm min}$ in
the published catalogue.

\section{325-MHz Source Counts}\label{sec:scounts}

\begin{figure}
\includegraphics[width=\linewidth]{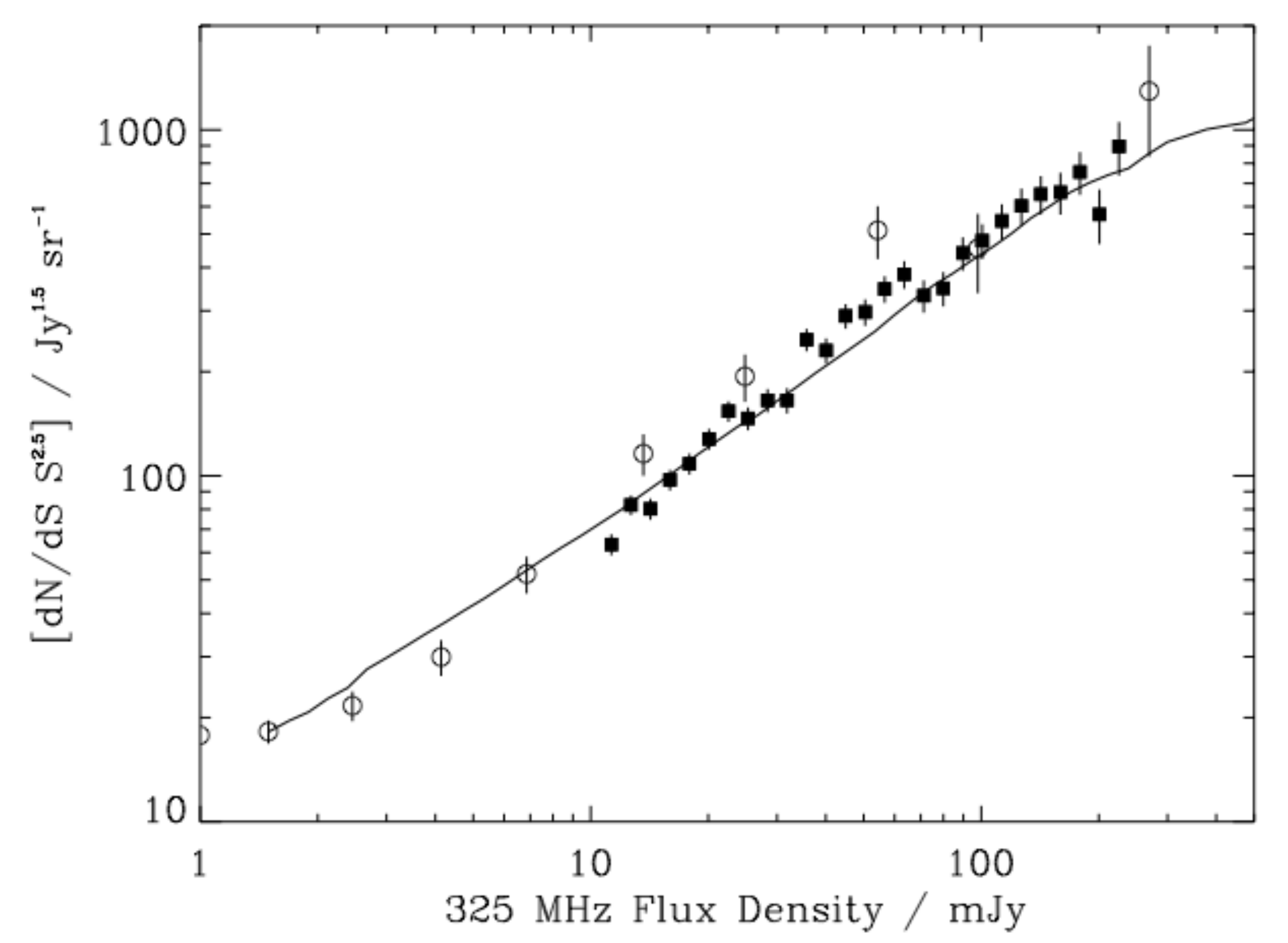}
\caption{The 325-MHz source counts measured from our GMRT survey
  (filled squares) and from the survey of the ELAIS-N1 field by
  \protect\cite{Sirothia2009} (open circles). The solid line shows the
  predicted source counts from the SKADS simulation \citep{Wilman08,Wilman10}.}
\label{fig:scounts}
\end{figure}

We have made the widest and deepest survey yet carried out at
325 MHz. It is therefore interesting to see if the behaviour of the
source counts at this frequency and flux-density limit differ from
extrapolations from other frequencies. We measure the source counts
from our GMRT observations using both the catalogues and the rms noise
map described in Section \ref{sec:noise}, such that the area available
to a source of a given flux-density and signal-to-noise ratio is
calculated on an individual basis. We did not attempt to merge
  individual, separate components of double or multiple sources into
  single sources in generating the source counts. However, we note
  that such sources are expected to contribute very little to the
  overall source counts. Fig.~\ref{fig:scounts} shows the source
counts from our GMRT survey compared to the source count prediction
from the Square Kilometre Array Design Study (SKADS) Semi-Empirical
Extragalactic (SEX) Simulated Sky \citep{Wilman08,Wilman10} and the
deep 325~MHz survey of the ELAIS-N1 field by \cite{Sirothia2009}. Our
source counts agree, within the uncertainties, with those measured by
\cite{Sirothia2009}, given the expected uncertainties associated with
cosmic variance over their relatively small field ($\sim
3$~degree$^{2}$), particularly at the bright end of the source counts.

The simulation provides flux densities down to nJy levels
at frequencies of 151 MHz, 610 MHz, 1400 MHz, 4860 MHz and
18 GHz. In order to generate the 325-MHz source counts from this
simulation we therefore calculate the power-law spectral index between
151 MHz and 610 MHz and thus determine the 325-MHz flux density. We
see that the observed source counts agree very well with the simulated
source counts from SKADS, although the observed source counts tend to
lie slightly above the simulated curve over the 10-200 mJy
flux-density range. This could be a sign that the spectral curvature
prescription implemented in the simulation may be reducing the flux
density at low radio frequencies in moderate redshift sources, where
there are very few constraints. In particular, the SKADS simulations
do not contain any steep-spectrum ($\alpha_{325}^{1400}<-0.8$)
sources, but there is clear evidence for such sources in the current
sample (see the following subsection). A full investigation of this is
beyond the scope of the current paper, but future observations with
LOFAR should be able to confirm or rebut this explanation: we might
expect the SKADS source count predictions for LOFAR to be slightly
underestimated.

\section{Spectral index distribution}
\label{sindexsec}

In this section we discuss the spectral index distribution of sources in the
survey by comparison with the 1.4-GHz NVSS. We do this both as
a check of the flux density scale of our GMRT survey (the flux density scale of
the NVSS is known to be better than 2 per cent: \citealt{nvss}) and as
an initial investigation into the properties of the faint 325-MHz radio
source population.

In all three fields the GMRT data have a smaller beam than the 45
arcsec resolution of the NVSS. We therefore crossmatched the two
surveys by taking all NVSS sources in the three H-ATLAS/GAMA fields and
summing the flux densities of the catalogued GMRT radio sources that
have positions within the area of the catalogued NVSS source (fitted
NVSS source sizes are provided in the `fitted' version of the
catalogue \citep{nvss}). 3951 NVSS radio sources in the fields
had at least one GMRT identification; of these, 3349 (85 per cent) of
them had a single GMRT match, and the remainder had multiple GMRT
matches. Of the 5263 GMRT radio sources in the survey 4746 (90 per
cent) are identified with NVSS radio sources. (Some of the remainder
may be spurious sources, but we expect there to be a population of
genuine steep-spectrum objects which are seen in our survey but not in
NVSS, particularly in the most sensitive areas of the survey, where
the catalogue flux limit approaches 3 mJy.)

\begin{figure}
\centering
\includegraphics[width=\linewidth]{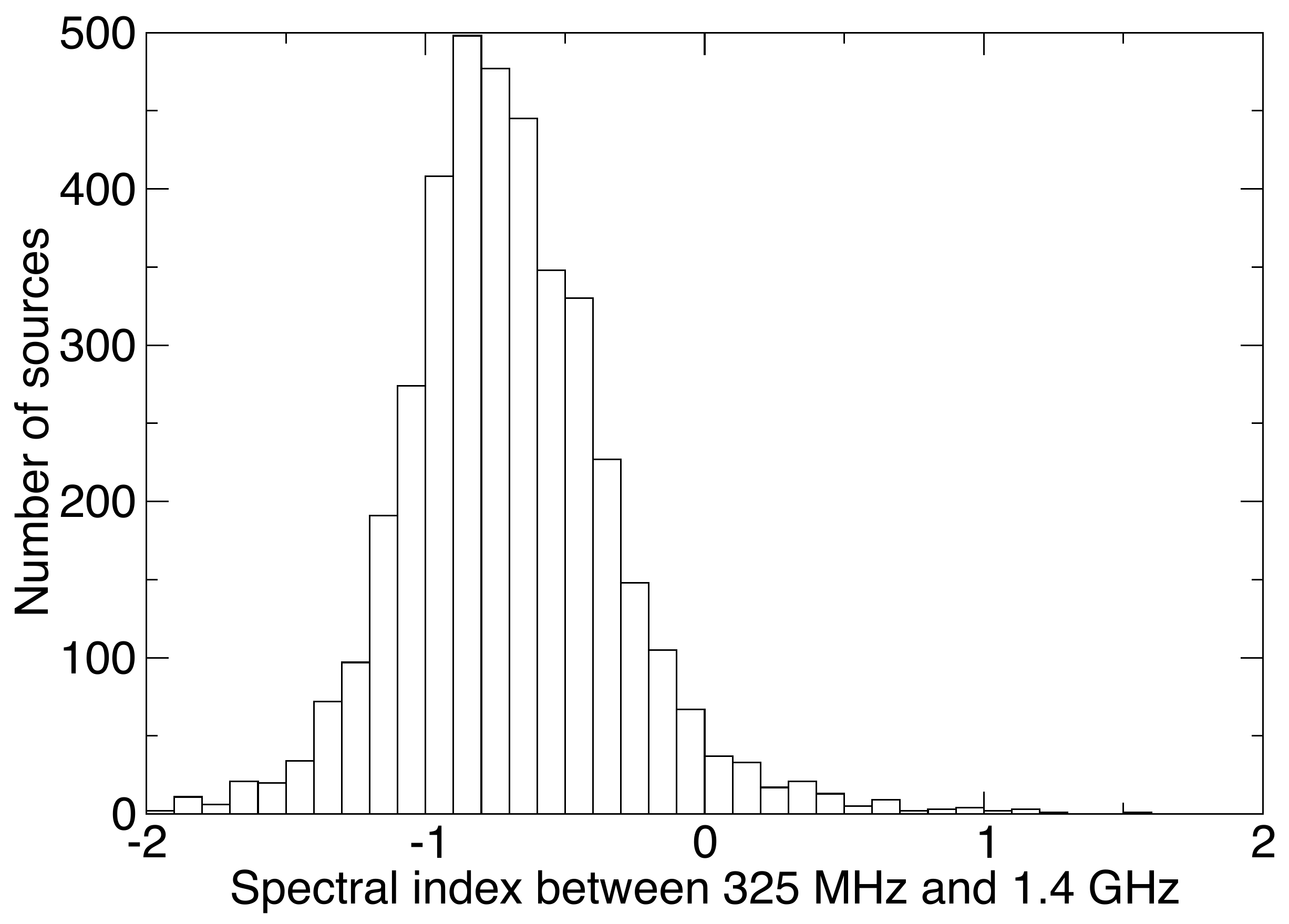}
\caption{The spectral index distribution between 1.4-GHz sources from the NVSS and 325-MHz 
GMRT sources.}
\label{sindex}
\end{figure}

Fig.~\ref{sindex} shows the measured spectral index distribution
($\alpha$ between 325 MHz and 1.4 GHz) of radio sources from the GMRT
survey that are also detected in the NVSS. The distribution has median
$\alpha=-0.71$ with an rms scatter of 0.38, which is in good agreement
with previously published values of spectral index at frequncies below
1.4 GHz \citep{sumss,debreuck2000,randall12}. (\cite{Sirothia2009}
find a steeper 325-MHz/1.4-GHz spectral index, with a mean value of
0.83, in their survey of the ELAIS-N1 field, but their low-frequency
flux limit is much deeper than ours, so that they probe a different
source population, and it is also possible that their use of FIRST
rather than NVSS biases their results towards steeper spectral
indices.) The rms of the spectral index distributions we obtain
increases with decreasing 325-MHz flux density; it increases from 0.36
at $S_{325}>50$ mJy to 0.4 at $S_{325}<15$ mJy. This reflects the
increasing uncertainty in flux density for fainter radio sources in
both the GMRT and NVSS data.

\begin{figure}
\centering
\includegraphics[width=\linewidth]{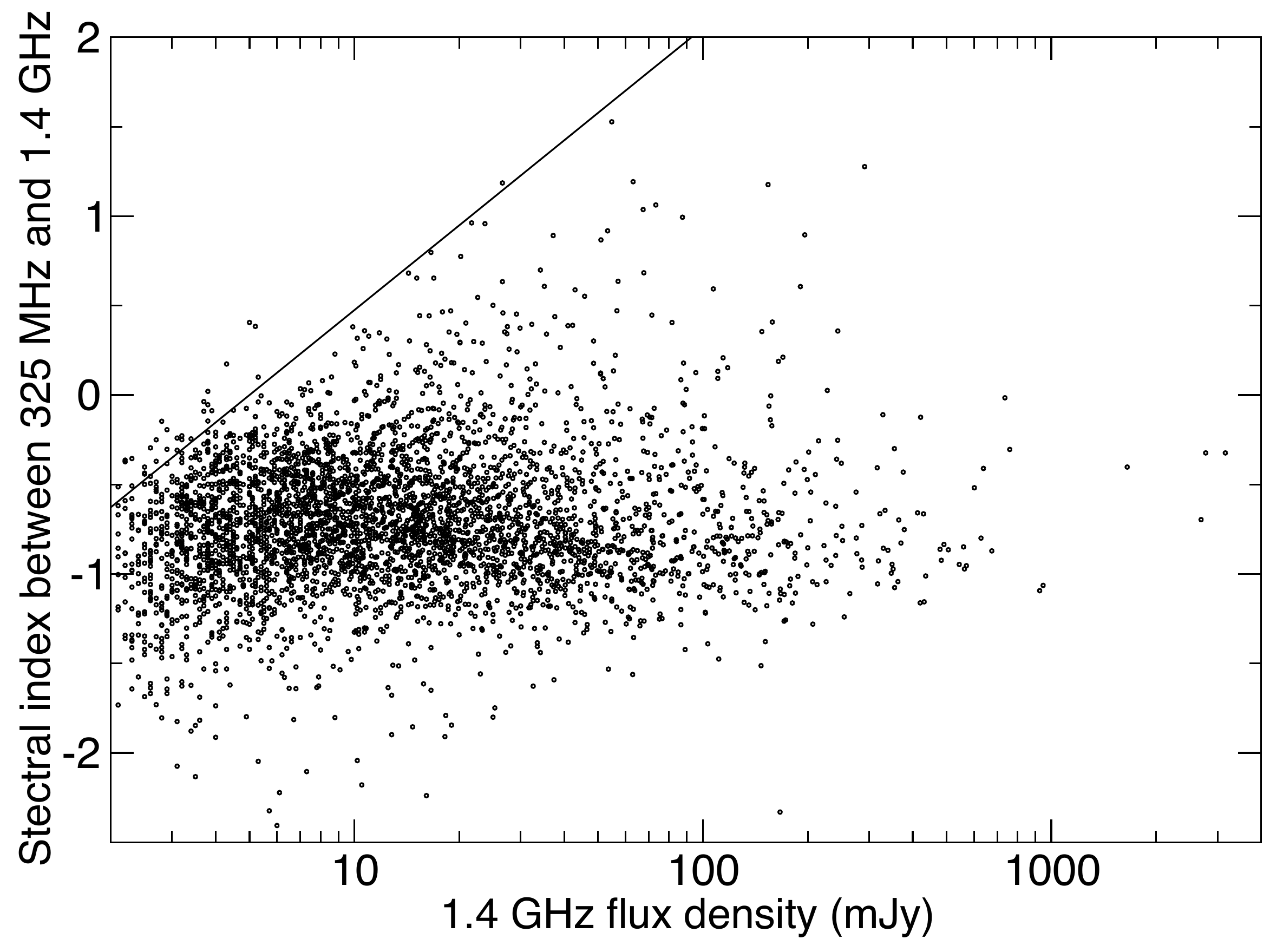}
\caption{The distribution of the spectral index measured between 325
  MHz and 1.4 GHz as a function of 1.4-GHz flux density. The solid
  line indicates the spectral index traced by the nominal 5 mJy limit
  of the 325-MHz data.}
\label{sindexflux}
\end{figure}

There has been some discussion about the spectral index distribution
of low-frequency radio sources, with some authors detecting a
flattening of the spectral index distribution below $S_{1.4}=10$ mJy
\citep{prandoni06,prandoni08,om08} and others not
\citep{randall12,ibar09}. It is well established that the 1.4-GHz
radio source population mix changes at around 1 mJy, with classical
radio-loud AGN dominating above this flux density and star-forming
galaxies and fainter radio-AGN dominating below it
\citep{condon+84,Windhorst+85}. In particular, the AGN population below 10 mJy
is known to be more flat-spectrum-core dominated
\citep[e.g.][]{nagar00} and it is therefore expected that some change
in the spectral-index distribution should be evident.
Fig.~\ref{sindexflux} shows the variation in 325-MHz to 1.4-GHz
spectral index as a function of 1.4-GHz flux density. Our data show
little to no variation in median spectral index below 10 mJy, in
agreement with the results of \cite{randall12}. The distribution
shows significant populations of steep ($\alpha < -1.3$) and flat
($\alpha > 0$) spectrum radio sources over the entire flux density
range, which are potentially interesting populations of radio sources
for further study (e.g.\ in searches for high-$z$ radio galaxies
\citep{hzrg} or flat-spectrum quasars).

\section{Summary}

In this paper we have described a 325-MHz radio survey made with the
GMRT covering the 3 equatorial fields centered at 9, 12 and 14.5-h
which form part of the sky coverage of {\it Herschel}-ATLAS. The data
were taken over the period Jan 2009 -- Jul 2010 and we have described
the pipeline process by which they were flagged, calibrated and
imaged.

The final data products comprise 212 images and a source catalogue
containing 5263 325-MHz radio sources. These data will be made
  available via the H-ATLAS (http://www.h-atlas.org/) and GAMA
  (http://www.gama-survey.org/) online databases. The basic data
  products are also available at http://gmrt-gama.extragalactic.info/
  .

The quality of the data varies
significantly over the three surveyed fields. The 9-h field data
has 14 arcsec resolution and reaches a depth of better than 1 mJy beam$^{-1}$
over most of the survey area, the 12-h field data has 15 arcsec
resolution and reaches a depth of $\sim 1$ mJy beam$^{-1}$ and the 14.5-h
data has 23.5 arcsec resolution and reaches a depth of
$\sim 1.5$ mJy beam$^{-1}$. Positions in the survey are usually better than
0.75 arcsec for brighter point sources, and the flux scale is
believed to be better than 5 per cent.

We show that the source counts are in good agreement with the
prediction from the SKADS Simulated Skies \citep{Wilman08,Wilman10}
although there is a tendency for the observed source counts to
slightly exceed the predicted counts between 10--100 mJy. This could
be a result of excessive curvature in the spectra of radio sources
implemented within the SKADS simulation.

We have investigated the spectral index distribution of the 325-MHz
radio sources by comparison with the 1.4-GHz NVSS survey. We find that
the measured spectral index distribution is in broad agreement with
previous determinations at frequencies below 1.4 GHz and find no
variation of median spectral index as a function of 1.4-GHz flux
density.

The data presented in this paper will complement the already extant
multi-wavelength data over the H-ATLAS/GAMA regions and will be made
publicly available. These data will thus facilitate detailed study of
the properties of sub-mm galaxies dectected at sub-GHz radio
frequencies in preparation for surveys by LOFAR and, in future, the SKA.

\section*{Acknowledgements}

We thank the staff of the GMRT, which made these observations
possible. We also
thank the referee Jim Condon, whose comments have helped to improve the
final version of this paper.
GMRT is run by the National Centre for Radio Astrophysics of the Tata
Institute of Fundamental Research. The {\it Herschel}-ATLAS is a project
with {\it Herschel}, which is an ESA space observatory with science
instruments provided by European-led Principal Investigator consortia
and with important participation from NASA. The H-ATLAS website is
http://www.h-atlas.org/. This work has made use of the University of
Hertfordshire Science and Technology Research Institute
high-performance computing facility (http://stri-cluster.herts.ac.uk/).

\setlength{\bibhang}{2.0em}
\setlength\labelwidth{0.0em}
\bibliography{allrefs,mn-jour}

\label{lastpage}
\end{document}